\documentclass[twocolumn,showpacs,preprintnumbers,amsmath,amssymb,showkeys]{revtex4}

\usepackage{graphicx}% Include figure files
\usepackage{dcolumn}% Align table columns on decimal point
\usepackage{bm}% bold math
\usepackage{amsmath}

\newcommand{\beq}{\begin{equation}}
\newcommand{\eeq}{\end{equation}}
\newcommand{\beqa}{\begin{eqnarray}}
\newcommand{\eeqa}{\end{eqnarray}}
\newcommand{\non}{\nonumber}

\begin{document}

\title{Spatial correlation functions in $3-d$ Ising spin glasses}

\author{Cirano De Dominicis}
\affiliation{SPhT, Orme des Merisiers, CEA-Saclay,
Gif-sur-Yvette, Cedex F-91191 (France)}
\email{cirano@spht.saclay.cea.fr}

\author{Irene Giardina}
\affiliation{
CNR ISC, Via dei Taurini 19, 00185 Roma (Italy)\\
and Dipartimento di Fisica\\
Universit\`a di Roma {\em La Sapienza}, P. A. Moro 2, 00185 Roma (Italy)}
\email{irene.giardina@roma1.infn.it}

\author{Enzo Marinari}
\affiliation{
Dipartimento di Fisica and INFN\\
Universit\`a di Roma {\em La Sapienza}, P. A. Moro 2, 00185 Roma (Italy)}
\email{enzo.marinari@roma1.infn.it}

\author{Olivier C. Martin}
\author{Francesco Zuliani}
\altaffiliation[Present address: ]{Nergal, Viale B. Bardanzellu 8, 
00155 Roma (Italy)}
\affiliation{LPTMS, Univ. Paris-Sud, Orsay Cedex F-91405 (France)}
\email{martino@ipno.in2p3.fr, zuliani@nergal.it}

\date{\today}

\begin{abstract}
We investigate spin-spin correlation functions in the low temperature phase
of spin-glasses. Using the replica field theory formalism, we examine in detail
their infrared (long distance) behavior.  In particular we
identify a {\em longitudinal mode}
that behaves as massive at intermediate length scales (near-infrared region).
These issues are then addressed by numerical simulation; 
the analysis of our data is compatible with
the prediction that the longitudinal mode appears as massive, i.e., that
it undergoes an {\it exponential} decay, for distances smaller than an
appropriate correlation length.

\end{abstract}


\maketitle

%%%%%%%%%%%%%%%%%%%%%%%%%%%%%%%%%%%%%%%%%%%%%%%%%%%%%%%%%%%%%%%%%%%%%%%
\section{\label{sec:motivations} Motivations}

Although spin glasses have been studied for over two decades, many
fundamental aspects of these systems~\cite{MezardParisi87b,Young98}
remain unclear.  In fact, some of the most basic questions relating to
their equilibrium properties remain open.  Roughly, there are
two schools of thought about spin-glass theories. In the
droplet/scaling picture~\cite{BrayMoore86,FisherHuse86}, at low
temperature $T$ there are just two equilibrium pure phases, related by
the up-down symmetry; there is no replica symmetry breaking (RSB) and
connected correlation functions decay to zero at long distance.  On
the contrary, in the mean-field picture
\cite{Parisi79,Parisi79b,Parisi80,Parisi80b,MezardParisi87b},
at low $T$ there are many pure states that are not related by the
up-down symmetry; because of this, generic connected correlation
functions {\em do not} decay to zero at large distances.

Much of the theoretical and numerical work on the low temperature
phase of spin glasses has been focused on the presence or
absence of replica symmetry breaking. The study of correlation
functions has not been pushed as much: at best, one of the spin-spin
correlation functions has been measured~\cite{MarinariParisi98} and
was found to decrease with a power law. Nevertheless, there are a
number of different theoretical predictions. Within the scaling and
droplet pictures~\cite{BrayMoore86,FisherHuse86,FisherHuse88}, 2-spin 
connected correlation functions generically decay to zero as $r^{-\theta}$,
where $r$ is the distance between the spins and $\nu\equiv -1/\theta$
is the usual thermal exponent.  For large dimensions $d$ one has
$\theta \approx \frac{d-1}{2}$ and numerical studies indicate that
$\theta(d=3)\approx 0.20$. 

The predictions of the mean-field picture have been more difficult to
obtain: it is necessary to go beyond the
Sherrington-Kirkpatrick~\cite{SherringtonKirkpatrick75} model because
in that system one cannot define distances.  Generally one does this
using replica field theory which allows computations
\cite{DominicisKondor97} in dimensions $d > 6$. Note that this
approach automatically forces one to consider disorder-averaged
quantities. Within this formalism, one finds that $2$-spin connected
correlation functions do not go to zero at large distances (because of
RSB) and that the large $r$ behavior is of the form $A + B\ r^{3-d}$.

In this work we reconsider these issues theoretically and numerically
in $d=3$. From replica field theory, we find the remarkable property
that a certain linear combination of correlation functions decays to
zero at large distance; furthermore,
this decay is ``fast'' in an
intermediate region, as corroborated by our numerical
measurements.

The outline of this paper is as follows.  In Section
\ref{sec:framework} we specify our model and we define the correlation
functions that we investigate.  In Section~\ref{sec:RFT} we discuss
the replica field theory approach and we compute the correlation
functions of interest in the tree approximation (no loops).  We then
move on to the numerical simulations. The corresponding methods are
presented in Section~\ref{sec:computational} while results are
presented in Section~\ref{sec:results}. We end with our conclusions.

%%%%%%%%%%%%%%%%%%%%%%%%%%%%%%%%%%%%%%%%%%%%%%%%%%%%%%%%%%%%%%%%%%%%%%%
\section{\label{sec:framework} General Framework}
%%%%%%%%%%%%%%%%%%%%%%%%%%%%%%%%%%%%%%%%%%%%%%%%%%%%%%%%%%%%%%%%%%%%%%%

%%%%%%%%%%%%%%%%%%%%%%%%%%%%%%%%%%%%%%%%%%%%%%%%%%%%%%%%%%%%%%%%%%%%%%%
\subsection{\label{subsec:models} The Model}
%%%%%%%%%%%%%%%%%%%%%%%%%%%%%%%%%%%%%%%%%%%%%%%%%%%%%%%%%%%%%%%%%%%%%%%

At a microscopic level, we consider a $d$-dimensional lattice of Ising
spins, $S_i=\pm1$, coupled by nearest-neighbor interactions.  The
corresponding Hamiltonian is
\begin{equation}
  H_J \equiv  -
  \sum_{\langle ij\rangle} J_{ij} S_i S_j - B \sum_i S_i\ ,
\end{equation}
where we have introduced an external magnetic field $B$. Ferromagnetic
couplings correspond to $J_{ij}>0$, anti-ferromagnetic ones to
$J_{ij}<0$. These couplings are quenched independent random variables
symmetrically distributed around $0$. To simplify the analytic
part of our study, it is most appropriate to take them to be Gaussian
of zero mean, while for our $d=3$ numerical study, we use $J_{ij}=\pm
1$ for simulational reasons (the difference is irrelevant in the
temperature region we explore). 

All the issues we consider in the following concern temperatures $T$
low enough so that one is within the frozen spin-glass
phase. Furthermore, we shall always take the $B\to 0$ limit. Within
the scaling/droplet picture, this limit leaves one with a single
equilibrium phase; on the contrary, in the mean-field picture, $B\to
0$ simply selects one partner for each pair of $B=0$ equilibrium
phases (because of RSB, there is an infinite number of these pairs).

%%%%%%%%%%%%%%%%%%%%%%%%%%%%%%%%%%%%%%%%%%%%%%%%%%%%%%%%%%%%%%%%%%%%%%%
\subsection{\label{subsec:Cfunctions} 2-Spin Correlation Functions}
%%%%%%%%%%%%%%%%%%%%%%%%%%%%%%%%%%%%%%%%%%%%%%%%%%%%%%%%%%%%%%%%%%%%%%%

For any given sample, consider the standard
spin-spin connected correlation function:
$$
  \langle S_i S_j \rangle - 
  \langle S_i \rangle \langle S_j \rangle\ .
$$
This quantity is not gauge invariant: it depends on
the (arbitrary) choices of $z$ axis orientations at sites $i$ and $j$.
To restore gauge invariance, one should consider
either the square of this correlation function
\begin{equation}
\label{eq:Gamma1}
\Gamma_1(i,j) = \lbrack \langle S_i S_j \rangle - 
\langle S_i \rangle \langle S_j \rangle \rbrack^2\ ,
\end{equation}
or the difference of the squares,
\begin{equation}
\label{eq:Gamma2}
\Gamma_2(i,j) = \langle S_i S_j \rangle^2
- \langle S_i \rangle^2 \langle S_j \rangle^2 \ .
\end{equation}
In all these quantities, $\langle \cdot \rangle$ refers to the
(thermal) ensemble average with respect to the usual Boltzmann
measure.  If $B=0$, by symmetry we have $\langle S_i \rangle=0$; that
is why we have introduced the infinitesimal magnetic field. Note that
$\langle \cdot \rangle$ does \emph{not} correspond to an average inside a
pure state; $\langle S_i \rangle^2$ is thus not the 
Edwards-Anderson~\cite{EdwardsAnderson75} order parameter.

In our study, we focus on disorder averages, averaging our observables
over samples where the couplings $J_{ij}$ are randomly
chosen with their a-priori distribution. We thus define
\begin{equation}
G_1(i,j) \equiv \overline{\Gamma_1(i,j)}  \ \ \ 
G_2(i,j) \equiv \overline{\Gamma_2(i,j)}\ ,
\label{eq:G1G2} 
\end{equation}
where the overline denotes this disorder average. 

What do we expect for these quantities?  Clearly, in the
droplet/scaling picture, the limit $B\to 0$ leads to a single
pure phase and so connected correlation functions go to zero.
Furthermore, the large distance behavior is controlled by a zero
temperature disordered fixed point.  Roughly, this leads to effective
couplings of ``block'' spins at large scales that have a broad
distribution: typical values scale as $r^{\theta}$ for blocks at
distance $r$. Since this distribution is bell shaped with a non-zero
value at the origin, droplets of characteristic scale $r$ are
thermally excited with a probability going as $k T r^{-\theta}$. This
feature leads to connected correlation functions decaying as $r^{-\theta}$.

In the mean-field picture the low $T$ phase undergoes RSB: connected
correlation functions do not go to zero at large $r$.  Interestingly,
the $r \to \infty$ limit of the disorder averaged $\langle S_i S_j
\rangle^2$ can be expressed simply in terms of a moment of
configurational overlaps. One defines the spin overlap $q$ of two
configurations $(1)$ and $(2)$ as
\begin{equation}
\label{eq:q12}
q^{(1,2)} \equiv \frac{\sum_i S_i^{(1)} S_i^{(2)}}{N}\ ,
\end{equation}
where $N$ is the total number of spins. A straightforward
computation of $\overline{\langle q^2 \rangle}$ gives
\begin{equation}
\label{eq:q2}
\overline{\langle q^2 \rangle} = 
\frac{\sum_{ij} \overline{\langle S_i S_j \rangle^2}}{N^2}
\end{equation}
The large volume limit of this quantity can be obtained by replacing
the correlation functions on the right hand side by their long
distance limit: in this way one sees that the long distance limit of
the first contribution to $G_2$ is $\overline{\langle q^2 \rangle}$.
As we shall see later, the mean-field formalism also predicts a very
non-trivial relation between the limiting values of $G_1$, $G_2$ and a
certain combination of moments of overlaps.

A more subtle question for the mean-field picture is how the
correlation functions approach their large $r$ limits. The framework
for addressing this question is replica field theory, which leads to
the conclusion~\cite{DominicisKondor97} that the decay should follow a
power law in $r$.   Our goal here is to
reconsider that computation, focusing in particular on linear
combinations of $G_1$ and $G_2$.

%%%%%%%%%%%%%%%%%%%%%%%%%%%%%%%%%%%%%%%%%%%%%%%%%%%%%%%%%%%%%%%%%%%%%%%
\section{\label{sec:RFT}Replica Field Theory for Generalized 
Correlation Functions}

%%%%%%%%%%%%%%%%%%%%%%%%%%%%%%%%%%%%%%%%%%%%%%%%%%%%%%%%%%%%%%%%%%%%%%%
\subsection{\label{subsec:starting}Starting Point}
%%%%%%%%%%%%%%%%%%%%%%%%%%%%%%%%%%%%%%%%%%%%%%%%%%%%%%%%%%%%%%%%%%%%%%%

We consider the linear combination of correlation functions $\alpha
G_1 + (1-\alpha) G_2$ which can be written as
\begin{eqnarray}
\nonumber
\chi^{\alpha}(i,j)&=& \overline{\langle S_i S_j\rangle^2} 
-2\alpha \overline
{\langle S_i S_j\rangle \langle S_i\rangle\langle S_j\rangle}
\\ 
\label{eq:corre}
&+& (2\alpha-1) \overline{\langle S_i\rangle^2 \langle S_j\rangle^2}\ , 
\end{eqnarray}
where $\alpha$ is a parameter that can take real values. The
brackets indicate thermodynamical averages in the presence of the
infinitesimal magnetic field, while the overline indicates an average
over the disorder distribution. In the case $\alpha=1$, after summing
over $j$,  we obtain the usual expression for the spin-glass
susceptibility $\chi$.

The starting point of replica field theory is the (truncated)
effective Lagrangian describing the replicated theory~\cite{DominicisKondor97}
\begin{eqnarray}
\nonumber
{\cal L} &\equiv&
-\frac{1}{4}\sum_{a,b}\sum_{\bf p}
(p^2-2\tau)\Phi^{ab}({\bf p})\Phi^{ab}({\bf -p})
\\ \label{eq:lagrangian}
&+&\frac{w}{6\sqrt{N}}\sum_{a,b,c}\sum_{\bf \{p_i\}}
\Phi^{ab}({\bf p_1})\Phi^{bc}({\bf p_2})
\Phi^{ca}({\bf p_3}) 
\\ \nonumber
&+&\frac{u}{12 N}\sum_{a,b}\sum_{\bf \{p_i\}}
\Phi^{ab}({\bf p_1})
\Phi^{ab}({\bf p_2})\Phi^{ab}({\bf p_3})
\Phi^{ab}({\bf p_4})\ ,
\end{eqnarray}
where all the sums over momenta are constrained in such a way that the
total momentum adds to zero.  Note that to reduce the complexity of
the presentation, we have used discrete sums over momenta rather than
the integrals that arise in the thermodynamic limit, $N \to \infty$.
Expanding around the mean-field solution
\begin{equation}
\Phi^{ab}({\bf p})=\delta_{{\bf p},0} \ q_{ab}+\phi^{ab}({\bf p})\ ,
\label{eq:sepa}
\end{equation}
one finds
\begin{eqnarray}
{\cal L}&=&{\cal L}_{mf}(q)  
-\frac{1}{2}\sum_{a<b,c<d}\sum_{\bf p}\phi^{ab}({\bf p}) 
M^{ab;cd}({\bf p})\ \phi^{cd}(-{\bf p})\nonumber \\ 
&+& {\cal L}_{int}(\phi) \ ,
\label{eq:lagrangian2}
\end{eqnarray}
where
\begin{equation}
\label{eq:lagrangianMF}
{\cal L}_{mf}(q)=N \{
\frac{\tau}{2}\sum_{ab}q_{ab}^2
+\frac{w}{6}\sum_{a,b,c}q_{ab}q_{bc}q_{ca}
+\frac{u}{12}\sum_{a,b}q_{ab}^4
\}\ ,
\end{equation}
$M$ is the matrix of the Gaussian fluctuations 
and ${\cal L}_{int}$ is an interaction term including
contributions up to the fourth order in $\phi$ (see
Eq.~(\ref{eq:lagrangian})).

In the context of this replicated field theory, the spin correlation
function that appears in Eq.~(\ref{eq:corre}) can be expressed in terms
of field correlations. For example, we have
\begin{equation}
\overline{\langle S_i S_j\rangle^2}
=\overline{\langle S_i^{(1)} S_i^{(2)}  S_j^{(1)} S_j^{(2)}\rangle}
=\langle \Phi^{12}_i\Phi^{12}_j \rangle_{\cal L}
=C^{12;12}(i,j)\ ,
\end{equation} 
where $\langle ...\rangle_{\cal L}$ is an average performed using the
effective Lagrangian of Eq.~(\ref{eq:lagrangian}) and where
the superscripts on the spins are replica indices. In the low
temperature phase, where one assumes that many states may exist, 
the expression for $\chi^{\alpha}(i,j)$ leads one
to consider all the possible ways in which $4$ replica indices
can appear. Then we have the following general
expression (in position space, or, equivalently in momentum space):
\begin{eqnarray}\nonumber
\chi^{\alpha}&=&\frac{1}{n(n-1)}\sum_{a,b}\{C^{ab;ab}-\frac{2\alpha}{(n-2)}
\sum_{c\neq b} C^{ab;ac} \\
&+&\frac{(2\alpha-1)}{(n-2)(n-3)}\sum_{c,d\neq a,b}
C^{ab;cd}\}\ 
\label{eq:corre2}
\end{eqnarray}
where we have dropped the $(i,j)$ dependence 
of $\chi^{\alpha}$ and of the $C$s to lighten the notation.

We can now separate, as in Eq.~(\ref{eq:sepa}), the fluctuating part from the 
mean-field part:
\begin{eqnarray}
\nonumber
\label{eq:C}
C^{ab;cd}(i-j)&=&\langle \Phi^{ab}_i \Phi^{cd}_j\rangle=
q_{ab}q_{cd}+\langle \phi^{ab}_i \phi^{cd}_j\rangle\non\\ \nonumber
&=&q_{ab}q_{cd}+ G^{ab;cd}(i-j)\\
&+&{\rm loop \ \ corrections}\ ,
\end{eqnarray}
where $G=M^{-1}$ is the free propagator (see Eq.~(\ref{eq:lagrangian2})),
and where we schematically denote
by $i-j$ the distance between $i$ and $j$.
From this we get
\begin{eqnarray} \nonumber
\chi^{\alpha}&=&\chi^{\alpha}_{mf}+\chi^{\alpha}_{fl}
=\frac{2-\alpha}{3n(n-1)}[\sum_{ab}q_{ab}^2+\sum_{abc}q_{ab}q_{ac}]\non\\
\nonumber
&+&\frac{1}{n(n-1)}\sum_{a,b}\{G^{ab;ab}-\frac{2\alpha}{(n-2)}
\sum_{c\neq b} G^{ab;ac}\\ \nonumber
&+&\frac{(2\alpha-1)}{(n-2)(n-3)}\sum_{c,d\neq a,b}
G^{ab;cd}\}\\
&+&{\rm loop \ \ corrections\ .}
\label{eq:corre3}
\end{eqnarray}
Note that in the mean-field contribution $\chi^\alpha_{mf}$ the sums 
run over all replica indices but we have included
only the terms which have a finite limit when $n\to 0$. 

We now compute the generalized correlation function
$\chi^{\alpha}$ considering the mean-field solution together with the
Gaussian fluctuations around it, disregarding the loop corrections in
Eq.~(\ref{eq:corre3}).

%%%%%%%%%%%%%%%%%%%%%%%%%%%%%%%%%%%%%%%%%%%%%%%%%%%%%%%%%%%%%%%%%%%%%%%%%%%%
\subsection{\label{subsec:meanfield}Mean-Field Solution and Free Propagators}

\begin{figure}
\includegraphics[width=5.5cm,angle=0]{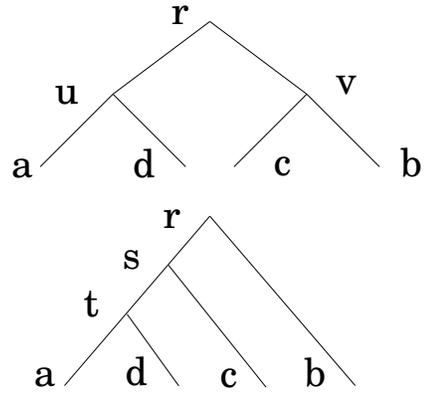}
\caption{Topologies associated with the replicon
sector (top) and the longitudinal anomalous sector (bottom).}
\label{fig:sectors}
\end{figure}

The mean-field solution $q_{ab}$ is obtained by minimizing the
Lagrangian ${\cal L}_{mf}$; it is thus given by the saddle point equation
\begin{equation}
\label{mf}
2\tau q_{ab}+w(q^2)_{ab}+\frac{2u}{3}q_{ab}^3=0\ .
\end{equation}
This equation is solved using an ultrametric Ansatz for the form of
the matrix $q$ ~\cite{Parisi79b,Parisi80,Parisi80b}. According to 
this Ansatz the matrix
$q$ is hierarchically divided into blocks 
by means of a procedure based on $R$ steps:
at each step $r$ the size of the blocks is set to $p_r$, with
$p_0=n$ and $p_{R+1}=1$.  If two replica indices belong to
the same block of size $p_r$ but to two distinct blocks of size
$p_{r+1}$ then $q_{ab}=q_r$, and we say that their ``co-distance''
$a\cap b$ in replica space is $r$. In this way the matrix $q$ is
parameterized by the $R$-step function $q_r$, where $r\in [0,R]$; note
that each $q_r$ has a multiplicity $\delta r=p_r-p_{r+1}$.  One can
also represent this procedure with blocks via a hierarchical tree, having
its root at $r=0$, its leaves at $r=R$, and having $p_{r}/p_{r+1}$
branches at nodes of height $r$. Replica indices lie on the leaves and their
overlap is $q_r$ if they join at a node of height $r$ on the tree.
At the end
one takes the limit $R\to\infty$ and $n\to0$: now the values $p_r$
tend to a monotonic increasing function that takes values in $[0,1]$. 
In this limit,
setting for example $p_r=x$ with $x\in [0,1]$, $q_r$ becomes a
continuous function $q(x)$
and  we indicate the replica co-distance as $a\cap b= x$.  
The well-known mean-field solution of Eq.~(\ref{mf}) then reads:
\begin{eqnarray}\nonumber
q(x)&=&\frac{w}{2u}x \quad \quad \quad \quad x<x_1\\ \nonumber
q(x)&=&q_1=\frac{w}{2u}x_1 \quad x_1<x<1\\
\rm{with} \quad \quad \tau&-&wq_1+uq_1^2=0\ .
\label{eq:sol1}
\end{eqnarray}
To calculate the free propagator $G$, one has to evaluate the 
fluctuation matrix 
\begin{equation}
M^{ab;cd}=\frac{\partial^2{\cal L}_{mf}(q)}
{\partial q_{ab}\partial q_{cd}}
\label{eq:fluct}
\end{equation}
given the mean-field solution Eq.~(\ref{eq:sol1}), and then 
invert it.  To do
this, it is convenient to parametrize both the fluctuation matrix $M$
and the propagator $G=M^{-1}$ in a way that
exploits the ultrametric structure of
the mean-field solution.  Each element of $M$ and $G$ is in principle
labeled by four replica indices. However, due to ultrametricity,
three replica co-distances are enough to characterize the geometrical
structure of the four indices. In terms of these co-distances we can then
identify two possible ``sectors'' for the propagator $G^{ab;cd}$ (and
for $M^{ab;cd}$):
\begin{itemize}
\item[i)] The replicon sector: when $a\cap b=c\cap d \equiv r$
\begin{equation}
G^{ab;cd}=G^{r,r}_{u,v}\ ,
\label{eq:propa-repli}
\end{equation}
with $u=\max(a\cap c, a\cap d)$,  $v=\max(b\cap c, b\cap d)$ and $u,v > r$.
\item[ii)] The longitudinal-anomalous sector:
when $a\cap b \equiv r$ and  $c\cap d \equiv s$
\begin{equation}
G^{ab;cd}=G^{r,s}_{t}\ ,
\label{eq:propa-la}
\end{equation}
with $t=\max(a\cap c,a\cap d,b\cap c,b\cap d)$.
\end{itemize}
These two sectors are illustrated in Fig.~\ref{fig:sectors}.

The inversion of the fluctuation matrix is not simple
\cite{DominicisKondor97}. It can be carried out using the Replica
Fourier transform when $d>6$, since in this case there are no
infra-red divergences (see~\cite{DominicisCarlucci97,CarlucciThesis}
and the Appendix \ref{app:appendix1} of this paper), enabling one to
block-diagonalize the fluctuation matrix. In this way the propagators $G$
are expressed in terms of Replica Fourier transformed variables, whose
explicit form in terms of the mean-field solution $q_{ab}$ can be
computed.  In the next section we will use these expressions to
compute the generalized correlation functions.

%%%%%%%%%%%%%%%%%%%%%%%%%%%%%%%%%%%%%%%%%%%%%%%%%%%%%%%%%%%%%%%%%%%%%%%%
\subsection{\label{subsec:generalized}The Generalized Correlation Functions}

The expression for the mean-field part $\chi^{\alpha}_{mf}$ is easily
obtained using Eq.~(\ref{eq:sol1}):
\begin{eqnarray}
\chi^{\alpha}_{mf}&=&\frac{2-\alpha}{3} 
\left [ \int_0^1dx\ q^2(x)
-\left(\int_0^1 dx\ q(x)\right)^2\right ]\non\\
&=& \frac{2-\alpha}{3}\frac{w^2}{4u^2}
\left [\frac{x_1^3}{3}-\frac{x_1^4}{4}\right ]\ .
\label{eq:constant}
\end{eqnarray}
Note that this expression gives for $\alpha=2$ a mean-field value
equal to zero. As we will discuss later, this fact is 
the consequence of a ``sum rule''. However, as we
shall see in the next sub-section, the case $\alpha=2$ is peculiar
also at the level of fluctuations.

In the Parisi \cite{MezardParisi87b} solution, $x(q)$, the inverse of
$q(x)$, 
has a simple expression in terms of $P(q)$:
$$
P(q)=\frac{dx(q)}{dq}\ ,
$$
where
$$
P(q)\equiv\overline{
\left\langle   
\delta
\left(  
q-\frac1N\sum_i S_i^{(1)} S_i^{(2)}
\right)
\right\rangle
} \ .
$$
This makes clear that
$$
\chi^\alpha_{mf} = \frac{2-\alpha}{3}
\left(
\overline{\left\langle q^2\right\rangle}
-
\overline{\left\langle q\right\rangle^2}
\right) \equiv \frac{2-\alpha}{3} \sigma_{|q|}\ .
$$
Note that in the present computation the overlaps are constrained to
positive values because of the infinitesimal positive magnetic field;
however, that will not be case for the numerical data which were
collected in zero field (see the next section). For this reason and
to use a common notation throughout the paper, we indicated the
variance of $q$ with respect to $P(q)$ in the presence
of an infinitesimal field as $\sigma_{|q|}$. 

For the fluctuating part, with the parameterization given in the
previous section, and exploiting the definition of Replica Fourier
transform, we obtain in the $R$-step Ansatz:
\begin{eqnarray}
\label{corre4}
&\chi^{\alpha}_{fl}&=\\ \nonumber
&-&\frac{2-\alpha}{3}\left \{\sum_0^R\delta r\sum_0^R\delta s \ G^{rs}_{R+1}
+\frac{1}{4}\sum_0^R\delta r \ G^{rr}_{\widehat{r+1}}\right.\\ 
\nonumber
&+&\left.\sum_0^R\delta r \ _RG^{rr}_{R+1,\widehat{r+1}}
+\sum_0^R\delta r \  G^{rr}_{R+1,R+1}\right \}\\ \nonumber
&-&\frac{2\alpha-1}{3}\left \{\sum_0^R\delta r \ 
G^{r,r}_{\widehat{r+1},\widehat{r+1}}+\frac{1}{2}
\sum_0^R\delta r \sum_0^R \delta s 
\ G^{r,s}_{\hat 0}\right \}\ ,
\end{eqnarray}
where the $\widehat{\cdots}$ indicates that we have used an operation
of Replica Fourier transform with respect to the hatted variable, and
$_RG^{r,r}_{u,v}$ stands for a pure replicon propagator
(where the longitudinal-anomalous 
contribution has been taken away, see Appendix \ref{app:appendix1}).

%%%%%%%%%%%%%%%%%%%%%%%%%%%%%%%%%%%%%%%%%%%%%%%%%%%%%%%%%%%%%%%%%%%%%
\subsubsection{\label{subsubsec:2}The Special Case $\alpha=2$}

In the case of $\alpha=2$ the first two lines of terms
in Eq.~(\ref{corre4}) 
disappear and the calculation simplifies a lot.
We start with the general expression for the replica
Fourier transform of the replicon propagator (see Appendix 
\ref{app:appendix1}):
\begin{equation}
G^{r,r}_{\hat{k},\hat{l}}=\frac{1}{p^2+\frac{w^2}{4u}[k^2+l^2-2r^2]}\ ,
\label{eq:RFT-repli}
\end{equation}
and thus the first sum gives 
\begin{eqnarray} \nonumber
\sum_0^R \delta r \  G^{r,r}_{\widehat{r+1},\widehat{r+1}}
&=&\sum_0^R (p_r-p_{r+1})
\frac{1}{p^2}\\ 
&=&(p_0-1)\frac{1}{p^2}\ \ \ 
\mathop{\longrightarrow}_{n\to 0}\ \ \  \frac{-1 ~}{p^2}\ .
\end{eqnarray}
The second sum is less straightforward. When we go to the continuous
limit, the variables $p_r,p_s$ with $r,s\leq R$ assume values in the
continuum interval $[0,x_1]$, while $p_{R+1} \equiv 1$; because of that, it
is convenient to separate the sum into two different pieces before taking
the limit of continuous replica symmetry breaking ($R\to\infty$):
\begin{eqnarray}\nonumber
&& \frac{1}{2}\sum_0^R \delta r\sum_0^R \delta s \  G^{r,s}_{\hat 0} = 
 \sum_{r\leq s\leq R-1} \delta r \ \delta s \ G^{r,s}_{\hat 0}\\ \nonumber
&+&(p_R-p_{R+1})\sum_{r\leq R-1}  \delta r
\  G^{r,R}_{\hat 0}+(p_R-p_{R+1})^2 \ G^{R,R}_{\hat 0}\\ \nonumber
&&\longrightarrow   \int_0^{x_1} dx
\int_x^{x_1}dy \  G^{x,y}_{\hat 0}\\ 
&+& (1-x_1)\int_0^{x_1} dx \  G^{x,x_1}_{\hat 0}
+(1-x_1)^2 \ G^{x_1,x_1}_{\hat 0}\ .
\label{sum}
\end{eqnarray}
Now we can use the expression of the longitudinal Fourier
transform obtained in~\cite{DominicisKondor97}:
\begin{eqnarray} \nonumber
G^{x,y}_{\hat 0}=\frac{2}{p^2\hat{p}}\ \ \frac{\sinh{\frac{x}{\hat p}}}
{\cosh{\frac{x_1}{\hat p}+\frac{1-x_1}{\hat p}\sinh{\frac{x_1}{\hat
p}}}}\\ 
\left [ \cosh{\frac{x_1-y}{\hat p}}+\frac{1-x_1}{\hat
p}\sinh{\frac{x_1-y}{\hat p}} \right ]\ ,
\label{eq:RFT-la}
\end{eqnarray}
where $\hat{p}^2=\frac{u}{w^2}\ p^2=\frac{x_1}{2wq_1} p^2$.
In this way we get for Eq.~(\ref{sum})
\begin{equation}
\frac{1}{p^2}\left [
1- {\hat p}\frac{\tanh{\frac{x_1}{\hat p}}+
\frac{1-x_1}{\hat p}} {1+\frac{1-x_1}{\hat p}\tanh{\frac{x_1}{\hat p}}} 
\right ]
\ ,
\end{equation}
and altogether we get
\begin{equation}
\chi_{fl}^{\alpha=2}(p)= \frac{\hat p}{p^2}\frac{\tanh{\frac{x_1}{\hat p}}+
\frac{1-x_1}{\hat p}} {1+\frac{1-x_1}{\hat p}\tanh{\frac{x_1}{\hat p}}} \ .
\end{equation}

Note  that the singularity present in the replicon part is canceled 
by the $1/p^2$ term in the longitudinal part, leaving a less divergent 
quantity in the long wavelength limit.
Let us now look more in details at the infrared behaviour of this correlation. 
In the  far-infrared region, i.e., for 
momenta much smaller than the small mass $2 x_1 w q_1$
(see~\cite{DominicisKondor97}), we have that $x_1/{\hat p} \to \infty$ and 
\begin{eqnarray}
&&\chi_{fl}^{\alpha=2}(p) \simeq \frac{x_1}{\hat p}
\left  (\frac{p^2}{2x_1wq_1})\right )^{1/2} \sim \frac{1}{p} \\ 
&& \mathrm{for}\ \ \ p^2 \ll 2 x_1 w q_1  \ . \nonumber
\label{farinfrared}
\end{eqnarray}
On the other hand, in the so called near-infrared region, i.e., in between
the small mass $2 x_1 w q_1$ and the large mass $2 w q_1$, we have that
$x_1/{\hat p} \ll 1$ and we get
\begin{eqnarray}
&&\chi_{fl}^{\alpha=2}(p) \simeq 
\frac{1}{p^2+2wq_1 (1-x_1)}\\
&& {\mathrm{for}}\ \ \ 2 x_1 w q_1 \ll p^2 \ll  2 x_1 w q_1 \ . \nonumber
\label{nearinfrared}
\end{eqnarray}
Thus, for intermediate momenta (or intermediate distances, i.e., between the
correlation lengths associated to the two masses), $\chi^{\alpha=2}_{fl}(p)$
behaves as if it were massive. Only at very large distance does one feel
the infrared singularity in $1/p$ (or the associated inverse power behaviour
in distance). The width of the region where the apparent massive behaviour
is felt depends on the size of $x_1$. For $6<d<8$, one has
$x_1\sim (w^2/\zeta)\ t^{(d-6)/2}$,
where $t$ is the reduced temperature $(T_c-T)/T_c$ and $\zeta$ the number 
of neighbours. As one reaches $d=6$ from above $x_1$ remains small like 
$w^2/\zeta$. Below $d=6$, $x_1$ is a function of $w^2/\zeta$ 
with, at the critical fixed point, $w^2/\zeta=\epsilon/2$, i.e.,
$x_1=f(\epsilon)\sim \epsilon/2$ for $\epsilon \ll 1$.
In three dimensions we have unfortunately no explicit 
knowledge of the size of $x_1$, though 
the previous computation shows that, if the two mass-scales remain well separated, there
exists a range of distances where massive-like behaviour occurs.

In the far-infrared region one can apply scaling rules to
guess what power-law behaviour arises below $d=6$. From the 
Kadanoff scaling hypothesis we expect the correlation function to decay as
$\chi^{\alpha=2}_{fl}(p) = \psi(p\ \xi) /p^{2-\eta}$, where 
$\xi \sim t^{-\nu}$ is the associated correlation length.  
From Eq.~(\ref{farinfrared}) we then get
\begin{equation}
\chi_{fl}^{\alpha=2}(p) \simeq \frac{x_1}{\hat p}
\left  (\frac{p^2}{2x_1wq_1}\right )^{1/2}
 \rightarrow\ \frac{x_1}{p^{2-\eta}}\left ( \frac{p^{\frac{1}{\nu}}}
{x_1 t}\right )^{\beta/2}
\end{equation}
that is
\begin{equation}
\chi_{fl}^{\alpha=2}(p) \sim
\frac{x_1}{p^{2-\eta}}\frac{p^{\frac{d-2+\eta}{4}}}{t^\beta}
\end{equation}
or by Fourier Transform
\begin{equation}
\chi_{fl}^{\alpha=2}(r) \sim
\frac{1}{r^{\frac{5}{4}(d-2+\eta)}} \ .
\end{equation}

Consider now instead the near-infrared regime. In this region, 
the generalized correlation function with $\alpha=2$ 
shows an exponential decay with distance, the asymptotic power
law decay arising only at still much larger length scales.
As we shall see, this behaviour is peculiar of $\alpha=2$. Indeed for every
other value of $\alpha$ the correlation function always decays 
algebraically (see Sec.~\ref{subsubsec:not2}).
Within the framework of the
scaling/droplet model~\cite{BrayMoore86,FisherHuse86,FisherHuse88}
correlation functions always decay algebraically, 
with no difference between near and far-infrared.

We note that in the near-infrared the $\alpha=2$ correlation function has 
a longitudinal nature in the usual field theory terminology. 
It is then appropriate to compare 
the present result with what happens in other systems where a continuous
symmetry is broken. It is known that in $O(N)$ systems the longitudinal mode,
which is massive at the level of Gaussian fluctuations, is converted into a 
massless mode when loop corrections are taken into account 
\cite{BrezinWallace73,BrezinWallace73b}, and one may wonder whether 
the same  also happens  in spin glasses.
The argument for the Heisenberg model, as 
detailed in Appendix (\ref{app:appendix2}), 
relies on the fact that the effective interaction 
between Goldstone modes vanishes in the infrared limit, and these modes 
are thus effectively free.  
In the Ising spin glass, however, the situation is 
different since in this case 
Goldstone modes remain coupled in the infrared \cite{DominicisKondor97}.
As a consequence, the argument for the $O(N)$ case does not apply here and 
the longitudinal mode $\chi^{\alpha=2}$ in the near-infrared 
is expected to remain massive: massive-like behaviour at 
intermediate scales is expected to occur
even if the loop corrections are included.

Finally, we note that a different behavior is found if our analysis for $d>6$ is extended
to configurations which are constrained to have a fixed
value of their overlaps. For example, for $r=s=0$, that is for configurations
having zero mutual overlap, the longitudinal part is identically zero,
and we are left with a singular $1/p^2$ behavior in the whole
infrared region.  (This happens in
fact for all values of $\alpha$, not just for $\alpha=2$.)

%%%%%%%%%%%%%%%%%%%%%%%%%%%%%%%%%%%%%%%%%%%%%%%%%%%%%%%%%%%%%%%%%%%%%%%%%%%%
\subsubsection{\label{subsubsec:not2}The Case $\alpha\neq 2$}

For $\alpha\neq 2$ a calculation similar to the one detailed in the
previous section enables us to estimate the most singular
contributions in the limit ${\bf p} \to 0$. We find that
\begin{equation}
  \chi_{fl}^{\alpha\neq 2}\sim \frac{{\rm A}+\log p}{p^3}
  +\frac{\rm B}{p^2}\ ,
\label{eq:alphagen}
\end{equation}
where $A$ and $B$ are two numerical coefficients. These expressions
are valid for $d>6$. Using general scaling arguments
for the integrals appearing in the calculation, 
at $d<6$ one can argue that Eq.~(\ref{eq:alphagen}) gives in real space
\begin{equation}
  \chi_{fl}^{\alpha\neq 2}\sim \frac{\rm A}{r^{\frac{3}{4} (d-2+\eta)}}
  +\frac{\rm B}{r^{d-2+\eta}}\ ,
\label{eq:twoExponents}
\end{equation}
where $\eta$ is an anomalous dimension. 
In $d=3$, two recent numerical estimates for the $J_{ij}=\pm 1$ model give 
$\eta = -0.35 \pm 0.05$~\cite{KawashimaYoung96}, 
and $\eta=-0.22\pm 0.02$~\cite{MariCampbell99}.
The corresponding values for the exponent 
$(3/4) (d-2+\eta)$ of the leading term of Eq.(\ref{eq:twoExponents}) are, respectively,
$0.49 \pm 0.04$ and $0.585 \pm 0.015$.

%%%%%%%%%%%%%%%%%%%%%%%%%%%%%%%%%%%%%%%%%%%%%%%%%%%%%%%%%%%%%%%%%%%%%%%%%%%%
\section{\label{sec:computational}Computational Methods}

%%%%%%%%%%%%%%%%%%%%%%%%%%%%%%%%%%%%%%%%%%%%%%%%%%%%%%%%%%%%%%%%%%%%%%%%%%%%
\subsection{\label{subsec:montecarlo}Monte Carlo Simulations}

It is interesting to test the analytical predictions we have
just provided (based on $d>6$
calculations) to what happens in a three dimensional 
system. This means estimating the
correlation functions of Eq.~(\ref{eq:G1G2}) from numerical
simulations. To do that, we have generated $512$ spin-glass
samples, where the couplings $J_{ij}$ were randomly set to $\pm 1$; we
considered mainly $12\times 12\times 12$ cubic lattices with periodic
boundary conditions (smaller lattices have also been used,
allowing us to see the qualitative behavior of
finite size effects).

For each such sample we produced over $1000$ equilibrium
spin configurations which were essentially independent (since
they were
separated by a large number of updates).  This was
achieved using the Parallel Tempering Monte Carlo
method~\cite{HukushimaNemoto96,ZulianiThesis} as follows. The 
temperatures used
went from $0.1$ to $2.0$ in steps of $\Delta T = 0.1$ (the critical
temperature $T_c$ in this model is about $1.1$).  At each temperature
we performed sweeps using the usual Metropolis algorithm and followed
this by trial exchanges between neighboring temperatures; we call this
sequence a \emph{pass} of the algorithm. The value 
of $\Delta T$ was such that these
exchanges succeeded with a probability close to $1/2$.

After thermalizing the system using a million such passes, we stored
configurations every 1000 passes for further analysis. With these
choices of parameters, we verified that the configurations kept were
compatible with being statistically independent by examining their
overlaps. Furthermore, we also checked that each configuration visited
many times all the temperature values as suggested
in~\cite{Marinari98} and that the distribution of overlaps for each
sample was symmetric with high accuracy.  Finally, the use of
bit-packing enabled us to speed up the computation (this trick is
particularly useful for the case of $J_{ij}=\pm1$ couplings). The
overall computation was performed on a cluster of Linux-based PCs
running at 333 MHz and represents the equivalent of $2$ years of
mono-processor time.

%%%%%%%%%%%%%%%%%%%%%%%%%%%%%%%%%%%%%%%%%%%%%%%%%%%%%%%%%%%%%%%%%%%%%%%%%%%%
\subsection{\label{subsec:decomposition}Configuration Space Decomposition}

All of the theory exposed in Section~\ref{sec:RFT} implies the
presence of an infinitesimal magnetic field. Performing this limit
numerically is a nuisance because of the necessity to have several
values of the magnetic field and to control the accuracy of the limit.
Fortunately, it is possible to avoid such an extrapolation by
remarking that the role of a very small magnetic field is to select
half of the $B=0$ configurations. The simplest implementation of this
is to keep only those configurations (generated at $B=0$) whose
total magnetization $M$ is positive. This should give the correct
selection when the lattice size $L$ tends to $\infty$, 
though of course with finite $L$ this
may lead to some undesirable effects.  What properties does one expect
in the thermodynamic limit?  In the mean-field picture, there are many
equilibrium states, all coming in pairs when $B=0$; if one adds an
infinitesimal field, only one state in each pair survives and the
overlap with the remaining states becomes positive.

In practice, we have found that the selection of those configurations
with $M>0$ left us with quite a few overlaps that were negative. (This
was severe for some disorder samples and less in others.) We thus
sought a better decomposition of the configuration space into two
parts related by the global spin flip symmetry. For each disorder
sample, we used the following procedure. (A nearly identical 
method was applied in~\cite{CilibertiMarinari04}).

In a first iteration, we successively consider the equilibrium
configurations, and flip them if their mean overlap with the previous
configurations is negative.  In the iterations thereafter, one repeats
these flips but now using the mean overlap with all the other
configurations.  In other words, we try to ``align'' the
configurations as much as possible; effectively, our procedure is a
kind of descent optimization method where we seek to have all mean
overlaps be positive; the first pass is useful for generating a good
starting set.  

In many cases, we find that this iterative procedure
converges to a fixed point where all mean overlaps are positive. This
procedure has then split the configuration space into two parts that
are related by the global spin flip symmetry. When the algorithm does
not converge, we repeat the procedure with a new first pass after
permuting randomly the configurations; this can lead to convergence,
but if it does not, we simply stop after some number of trials and
accept an ``imperfect'' decomposition.

Given the set of aligned configurations (about half of them
have been flipped), we can then look at individual
rather than mean overlaps. The main test of whether the
configuration space decomposition is ``perfect''
is then the positivity of \emph{all} the overlaps between
these aligned configurations.

%%%%%%%%%%%%%%%%%%%%%%%%%%%%%%%%%%%%%%%%%%%%%%%%%%%%%%%%%%%%%%%%%%%%%%%%%%%%
\subsection{\label{subsec:parametrization}
Parametrization of Correlation Functions}

Within the theoretical analysis, one controls only the small
momentum behavior of $\chi^{\alpha}$ (e.g., $p \to 0$ in 
formula~(\ref{eq:alphagen})), that is the
large distance region. There is thus some arbitrariness
in the choice of fitting functions when comparing the
data to the theoretical predictions. 
Our choice of functions is as follows.
Start with a correlation function $G(p)$ on the infinite
lattice that has a power law behavior $p^{-c}$ as $p \to 0$;
the simplest choice is
\begin{equation}
\label{eq:latticeG}
G({\bf p}) = \frac{1}{\left[ 2 \sum_{\mu} (1-\cos(p_{\mu}) \right]^{c/2}}
\end{equation}
as motivated by the requirement that $G({\bf p})$ be periodic
in the components $p_{\mu}$ of ${\bf p}$. (We work
in units of the lattice spacing.) This correlation function
can be used also on a finite size lattice of
size $L \times L \times L$ with periodic boundary conditions
as long as we take the correct discrete values of the $p_{\mu}$.
We thus take this $G({\bf p})$ and Fourier transform it
to coordinate space. All our measurements are for
$i$ and $j$ on an axis of the lattice, and when performing
the transform we must remove the ${\bf p}={0}$ 
contribution; this defines a function hereafter called
$G_c(r)$. An analogous procedure was used in~\cite{MarinariMonasson95}.
One can also perform this computation
with a mass term, leading to exponentially decaying
correlation functions; we do this by adding $m^2$
in the denominator of the right hand side of Eq.~(\ref{eq:latticeG})
and setting $c=2$, leading to the function $G_m(r)$. 
(We performed checks on our code using the high precision
values for such functions in~\cite{FriedbergMartin87}.)
The resulting family of functions we use for our fits are 
then $a + b G_c(r)$ and $a + b G_m(r)$.

%%%%%%%%%%%%%%%%%%%%%%%%%%%%%%%%%%%%%%%%%%%%%%%%%%%%%%%%%%%%%%%%%%%%%%%%%%%%
\section{\label{sec:results}Results}

%%%%%%%%%%%%%%%%%%%%%%%%%%%%%%%%%%%%%%%%%%%%%%%%%%%%%%%%%%%%%%%%%%%%%%%%%%%%
\subsection{\label{subsec:overlaps}Effect on Overlaps of the Decomposition}

In the thermodynamic limit, we expect configuration space
decomposition schemes to lead to aligned configurations
for which the overlaps are positive. However, one can
expect some schemes to work better than others when $L$ is 
relatively modest (in our case $L=12$). As we already
pointed out, the simplest procedure does not work so well,
so let us examine the overlaps found with our ``improved''
scheme previously described. For each sample,
we have viewed the distribution $P(q)$ of overlaps found
for each pair of configurations, both before and after
the alignment. In the majority of the samples,
the procedure leads to nearly all overlaps
being positive; yet there are some samples for which
one still has a significant number of overlaps
that are negative. 
\begin{figure}
\includegraphics[width=5.5cm,angle=-90]{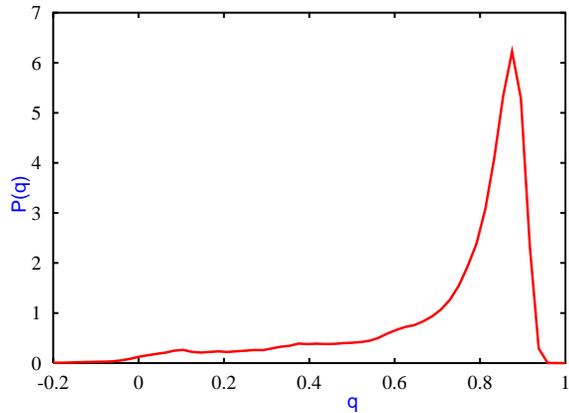}
\caption{Example of distribution of overlaps for a
disorder sample after the configuration space decomposition. $T=0.5$.}
\label{fig:pofq}
\end{figure}
In Fig.~\ref{fig:pofq}
we show an example where a rather
small fraction of the overlaps are negative. When considering
all 512 of our samples, the 
decomposition is reasonably good but not perfect;
this is not surprising considering our lattice size ($L=12$).
Because of these effects, we expect
our analysis of correlation functions 
to suffer from small systematic effects
since quantities
such as $\langle S_i \rangle$ are probably estimated 
with a residual bias.

%%%%%%%%%%%%%%%%%%%%%%%%%%%%%%%%%%%%%%%%%%%%%%%%%%%%%%%%%%%%%%%%%%%%%%%%%%%%
\subsection{\label{subsec:correlation}Correlation Functions}

Now we confront the different
predictions of the replica field theory approach to the
properties of the system as extracted from our simulations.

The first important prediction of replica field theory is that the
large distance limit of $\chi^{\alpha}$ is $(2-\alpha)/3$ times the
variance of $q$ in the presence of an infinitesimal magnetic
field. This prediction arises from the mean-field computation; it 
continues to hold when taking into account the 
Gaussian fluctuations; furthermore, as argued in the previous 
section, it should continue to hold within the loop expansion.
\begin{figure}
\includegraphics[width=6cm,angle=-90]{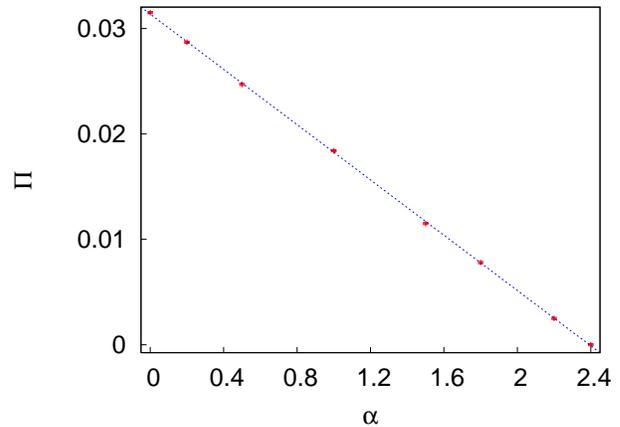}
\caption{Large distance value $\Pi$ of the correlation function
$\chi^{\alpha}$ of Eq.~(\ref{eq:corre}) as a function
of $\alpha$. Theoretically, the change of sign should be at 
$\alpha = 2$. (The data are for $T=0.5$.)}
\label{fig:plateau}
\end{figure}
We have fitted $\chi^{\alpha}(i,j)$ using the class of functions $a + b
G_c(r)$ and $a + b G_m(r)$ defined before.  We find that for $\alpha$
not close to $2$ the power fit is far superior and gives a good value
for $\chi^2$: there is at least one order of magnitude difference
between the power law best fit $\chi^2$ and the exponential best fit
$\chi^2$.  On the contrary, when $\alpha \approx 2$, the exponential
fit is better: in this case the decay is indeed so fast that only at
distance $1$ do we get sometimes a signal significantly different from
zero over the plateau value.

The constant value obtained from these fits gives the long distance
limit of $\chi^{\alpha}$; for $T=0.5$, we plot in Fig.~\ref{fig:plateau}
these limits as a function of $\alpha$. The behavior is linear
in $\alpha$ as it should, and the values change sign not far from
$\alpha=2$. Furthermore, when plotting the data as a function of $(2
-\alpha)/3$, the slope should be ${\overline{\langle q^2 \rangle}} -
\overline{\langle q \rangle} \overline{\langle q \rangle}$. The
numerical value we find for this slope is $0.040$; this has to be
compared to the numerically determined value of the variance of $|q|$
among our equilibrium configurations which is $0.047$ at
$T=0.5$: these two values are numerically close. Furthermore, 
since we find that this
variance decreases as $L$ increases, the two quantities probably 
do become equal at large $L$. Our measurements thus support
the theoretical claim that the large distance limit of
$\chi^{\alpha}$ is exactly given by Eq.~(\ref{eq:constant})
and are consistent with a large body of published results \cite{Young98}. 
The situation is similar at other temperatures in the low $T$ phase.
Notice that the plateau value at $\alpha=2$ is different from zero:
the effect is small, and on the $L=12$ data we get a zero plateau
close to $\alpha=2.4$. But this discrepancy decreases with
increasing lattice sizes, making manifest its finite size nature.

\begin{figure}
\includegraphics[width=6cm,angle=-90]{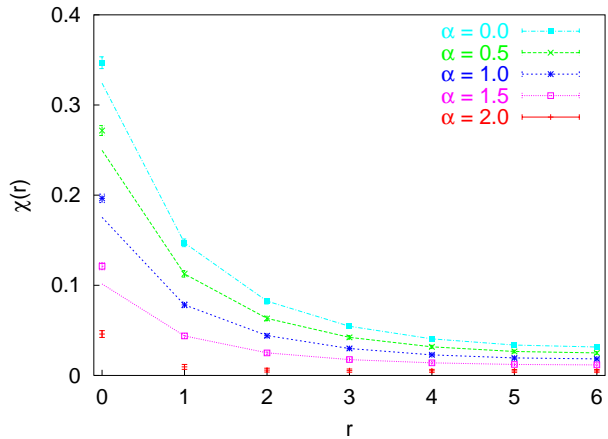}
\caption{Data for $\chi^{\alpha}(r)$ for 
$\alpha=0$, \ldots, 2 in steps of $0.5$ (from top to bottom). For
$\alpha\ne 2$ we display the best fits using $a + b G_c(r)$. 
(The data are for $T=0.5$.)}
\label{fig:gofr}
\end{figure}

Let us move on to the next prediction of the analytic computation
concerning \emph{how} the correlation functions tend toward their
large $r$ limit. In Fig.~\ref{fig:gofr} we show the raw data for
$\chi^{\alpha}(r)$ for $\alpha=0$, \ldots, 2 in steps of $0.5$. We
have also displayed the fits obtained when using the functional form
$a + b G_c(r)$ for the data $1 \le r \le 6$. For these values of
$\alpha$, the best values of the power $c$ are $2.51$, $2.49$, $2.45$
and $2.34$ for the first four.  These fits are good, in line with the
absence of visible systematic errors.  On the contrary, when using the
exponential fits, i.e., $a + b G_m(r)$, the fits are not very good
when $\alpha$ is away from $2$; in fact, the $\chi^2$ is minimized
when the mass tends to zero for $\alpha<1.5$. Finally, when
$\alpha=2$, the exponential fit is good and leads to a mass of $1.1$.

%
% here best fits are 
%  alpha 0.0 power 2.51
%  alpha 0.5 power 2.49
%  alpha 1.0 power 2.45
%  alpha 1.5 power 2.34
%  alpha 2.0 no best fit

Put all together, these 
analyses strongly favor a power law decrease
with $r$ when $\alpha \ne 2$ with a power that 
is consistent with being
$\alpha$-independent. To leading orders one then has
\begin{equation}
\chi^{\alpha\ne 2}(r) \approx \frac{2-\alpha}{3} \sigma_{|q|}^2
+ \frac{B}{r^{\lambda}}
\end{equation}
where $\sigma_{|q|}^2$ is the variance of $|q|$. In our data,
$\lambda=d-c$ is compatible with being an $\alpha$-independent
constant whose value is close to $0.5$.  This value may be compared
to the value $\theta=0.20$ expected within the scaling/droplet
picture, and to the exponent $(3/4) (d-2+\eta)$
predicted using scaling arguments within the
replica field theory calculation (equal to $0.49$ or $0.585$, according to the
numerical estimate of $\eta$ used in the formula - see Sec.~\ref{subsubsec:not2}).
In contrast, when $\alpha=2$, our data favor instead the fit
to an exponential decrease with $r$, and thus supports
the extrapolation of the $d>6$ theoretical
analysis of the near-infrared behaviour down to $d=3$.

%%%%%%%%%%%%%%%%%%%%%%%%%%%%%%%%%%%%%%%%%%%%%%%%%%%%%%%%%%%%%%%%%%%%%%%%%%%%
\section{\label{sec:discussion}Discussion}

We have investigated the properties of certain 
correlation functions in spin glasses
involving two spins. As a first result, we have found
that the large distance behavior of the two connected
correlation functions $G_1$ and $G_2$ (see Eq.~(\ref{eq:G1G2}))
satisfies 
\begin{equation}
\alpha G_1(\infty) + (1-\alpha) G_2(\infty) = 
\frac{2 - \alpha}{3} ~~ \lbrack \overline{\langle q^2 \rangle}
- \overline{\langle q \rangle}^2 \rbrack
\label{eq:GsInfty}
\end{equation}
and thus $G_1(\infty) = \sigma_{|q|}^2 / 3$ and
$G_2(\infty) = 2 \sigma_{|q|}^2 / 3$ where $\sigma_{|q|}^2$ is
the variance of the absolute value of the overlap, $|q|$.
(In all our computations, we assumed the presence
of an infinitesimal field:
the net effect of that field is to decompose the
configuration space, rendering overlaps positive.)

A second series of results concerns the way the correlation functions
$G_1$ and $G_2$ tend toward their limits.  We found numerically
a decrease close
to $1/r^{1/2}$ which is a bit different from the droplet/scaling prediction
of $1/r^{\theta}$ since $\theta \approx 0.2$. This power law
decay disappears when considering precisely $\alpha=2$, i.e., $2
G_1(r) - G_2(r)$: in that case one has both a limiting value compatible 
with zero and a very fast decay, suggestive of an exponential law.

All of our analytical predictions are based on replica field theory
calculations using an effective Lagrangian; this approach should be
reliable in dimensions $d > 6$. Nevertheless, 
the main predictions, namely the limiting values of $G_1$ and $G_2$ and
their approach to their limits, 
all are corroborated in $d=3$ by our numerical study.
Also the prediction  of a massive-like regime for $2 G_1 - G_2$ seems to be
nicely consistent with the numerical findings that show for this 
correlation a very fast decay.

It is of interest to note that Fisher and 
Huse \cite{FisherHuse86,FisherHuse88}
and Bray and Moore \cite{BrayMoore86} had also suggested that there
could be a peculiar behaviour for a longitudinal-like linear combination of
$G_1$ and $G_2$, due to the cancellation of the leading term. 
More precisely, the claim there was that the power 
law behaviour  would be
changed from $\theta\approx 0.2$ to a larger value (determined by the
expansion of the zero temperature distribution of the internal fields
around $h=0$ - see Sec. 4.1 of \cite{BrayMoore86}).  
We do not find a
decrease compatible with $1/r^{\theta}$  for $G_1$ and
$G_2$ separately, while we find a decay compatible with an exponential
law for $2 G_1 - G_2$, at least in the near infrared; thus 
qualitatively, the correspondence with the droplet is not good.

Coming back to our results for the large distance
limiting values of $G_1$ and $G_2$, we can derive
a simple ``sum rule'' as follows.
Each term in $\chi^{\alpha}(r)$ can be computed when $|i-j| \to \infty$
via replica moments just as we did in Eq.~(\ref{eq:q2}):
\begin{equation}
\lim_{r\to \infty} \overline{\langle S_i S_j \rangle  
\langle S_i \rangle  \langle S_j \rangle}  = \frac{1}{2}
[\overline{\langle q^2\rangle}+\overline{\langle q\rangle^2}]= 
\overline{\langle q_{12} q_{13}\rangle} \ ,
\label{eq:g2}
\end{equation}
and
\begin{equation}
\lim_{r\to \infty} \overline{\langle S_i \rangle^2  \langle S_j \rangle^2} =
\frac{1}{3}[2\overline{\langle q^2\rangle}+\overline{\langle q\rangle^2}] =
\overline{\langle q_{12} q_{34}\rangle}\ , 
\label{eq:g3}
\end{equation}
where, in the last equalities, the averages $\langle \cdots\rangle$ are 
performed with respect to a replicated equilibrium measure.
We then obtain the following sum rule when using
Eq.~\ref{eq:GsInfty} at $\alpha=2$:
\begin{equation}
\label{eq:sumrule1}
\overline{\langle q_{12}^2 \rangle} - 
4\overline{\langle q_{12}q_{13}\rangle} +3 
\overline{\langle  q_{12}q_{34} \rangle} =0  \ .
\end{equation}
It is also possible to derive this relation using
arguments based on stochastic stability~\cite{Guerra96}.

As a final comment, note that within the scaling or droplet pictures,
when the distance between $i$ and $j$ diverges one has
\begin{equation}
\lim_{|i-j| \to \infty}
\frac{ \overline{\langle S_i \rangle^2 \langle S_j \rangle^2} }
{\overline{\langle S_i \rangle^2} ~~ \overline{\langle S_j \rangle^2} } = 1 \ .
\end{equation}
This does not hold in the presence of replica symmetry breaking,
and instead we have a relation following 
from Eq.~(\ref{eq:g3}).

Several questions remain open.  One would like to determine
analytically the power law decrease of $G_1$ and $G_2$ which here was
compatible with $1/r^{1/2}$; is that the exact value, and what is it
in higher dimensions? Another issue is related to the presence of the
massive-like behaviour for the $\alpha=2$ correlation function in the 
near-infrared. It would be important to get some analytical 
estimates of the range where this behaviour does hold also in $d=3$. 
Finally, one may wonder whether there is any deep
reason for the cancellation of the leading  parts of $G_1$ and $G_2$
for $\alpha=2$. One may imagine that it is somehow related
to sum rules. Then one may conjecture that to each sum rule for
overlaps (obtained for instance using stochastic stability), there is
an associated correlation function which decays to zero anomalously
fast. This conjecture should be testable for several sum rules using Monte Carlo
methods.

%%%%%%%%%%%%%%%%%%%%%%%%%%%%%%%%%%%%%%%%%%%%%%%%%%%%%%%%%%%%%%%%%%%%%%%
\begin{acknowledgments}
We thank A.J. Bray and M.A. Moore for stimulating discussions,
and G. Biroli and J.-P. Bouchaud for their insights. 
I.G. thanks the Service de Physique Th\'eorique, CEA-Saclay, where
part of this work was done.
This work was supported in part by the European Community's
Human Potential Programme contracts
HPRN-CT-2002-00307 (DYGLAGEMEM) and
HPRN-CT-2002-00319 (STIPCO).
Furthermore, F.Z. was supported by an EEC Marie Curie fellowship
(contract HPMFCT-2000-00553).  The LPTMS is an Unit\'e de Recherche
Mixte de l'Universit\'e Paris~XI associ\'ee au CNRS.
\end{acknowledgments}

\appendix

%%%%%%%%%%%%%%%%%%%%%%%%%%%%%%%%%%%%%%%%%%%%%%%%%%%%%%%%%%%%%%%%%%
\section{\label{app:appendix1}Replica Fourier Transform}

The Replica Fourier transform is a powerful technique which greatly
simplifies the diagonalization and inversion of the fluctuation matrix
$G$ (see Eq.~(\ref{eq:lagrangian2}))). The idea is the following.
Convolutions in configuration space become products in Fourier space
when appropriately defining a Fourier transform in replica space;
thus sums over replica co-distances may be transformed into
associated products.  Similarly, matrices
can be inverted in the replica Fourier space and then
transformed back to get the desired expression.

Let us assume that replica objects (matrices, tensors etc.) only
depend on replica co-distances, and belong to the hierarchical structure
described in Sec. \ref{sec:RFT}. The Replica Fourier Transform (RFT)
\cite{DominicisCarlucci97} is a discretized and generalized version of the algebra
introduced in \cite{MezardParisi91}. For objects like $q_{ab}$ which
depend on a single replica co-distance $a\cap b=t$ the RFT ${q}_{\hat k}$
is defined as
\begin{equation}
{q}_{\hat k}=\sum_{t=k}^{R+1} p_t (q_t - q_{t-1}) \ , 
\label{eq:RFT}
\end{equation}
where the $p_t$ are the sizes of the Parisi blocks, and objects 
with indices out of the range of definition are taken as equal to zero.

The inverse transform is given by:
\begin{equation}
q_t= \sum_{k=0}^{t} \frac{1}{p_{k}}({q}_{\hat k}-{q}_{\widehat {k+1}}) \ .
\label{eq:RFT2}
\end{equation}
With these definitions, the convolution $\sum_c A_{ac} B_{cb} =C_{ab}$ 
becomes in RFT ${A}_{\hat k} {B}_{\hat k}={C}_{\hat k}$. 
However, in our case the problem is slightly more complicated. Starting from 
the known expression of the fluctuation matrix $M$ of Eq.~(\ref{eq:fluct}), 
we want to compute the propagator $G=M^{-1}$. 
Since both $M$ and $G$ are tensors bearing four indices, the unitarity  
equation $\sum_{cd} M^{ab;cd} G^{cd;ef}= \delta_{ab;ef}$ 
involves  a double  convolution  in replica space.  If the fluctuation matrix
and the propagator are expressed in terms of replica co-distances, as 
described in Eqs.~(\ref{eq:propa-repli}) and (\ref{eq:propa-la}), 
one needs to Replica Fourier transform with respect 
to lower indices (i.e., the indices referring to the  cross-overlaps between 
pairs  of replicas). 

In the Replicon sector, the double transform reads:
\begin{eqnarray}
\label{eq:RFT3}
M_{\hat k,\hat l}&=&
 \sum_{u=k}^{R+1} p_u \sum_{v=l}^{R+1} p_v 
\\ \nonumber
&&\left(
M^{r,r}_{u,v} - M_{u-1,v}^{r,r}
-M^{r,r}_{u,v-1}+ M^{r,r}_{u-1,v-1}\right ) \ ,
\end{eqnarray}
with, by definition, $k,l \ge r+1$. The inverse transformation can be
obtained by applying twice the inverse RFT defined in
Eq.~(\ref{eq:RFT2}). From the Lagrangian, Eq.~(\ref{eq:lagrangian}), we
can easily recover the explicit expression of the fluctuation matrix.
One has (with the notation of Sec.~\ref{subsec:meanfield}):
\begin{eqnarray}
M_{R+1,R+1}^{r,r} &=& p^2 - 2 (\tau + u q_r^2) \ ,\nonumber \\
M_{u,R+1}^{r,r} &=& -w q_u\ ,   \nonumber \\
M_{R+1,v}^{r,r} &=& - w q_v  \ . 
\label{eq:fluctu-repli}
\end{eqnarray}
From this, by noting that the mean-field solution gives in Replica 
Fourier space
$\tau= -w q_{\hat 0}$, one gets:
\begin{equation}
M^{r,r}_{\hat k,\hat l}= p^2 + w  [ (q_{\hat 0}-q_{\hat k}) + 
(q_{\hat 0} -q_{\hat l}) ]-2 u q^2_r \ .
\label{eq:RFT4}
\end{equation}
In the limit of an infinite number of steps of replica 
symmetry breaking, $q_r \to q(r)$, 
where now 
$r\in[0,1]$ while $q(r)$ obeys Eq.~(\ref{eq:sol1}). 
Correspondingly, the Replica Fourier transform becomes
$ q_{\hat k} = \int_k^{x_1} dx x dq/dx - 2 u q(x_1)$ and  
\begin{equation}
M_{\hat k,\hat l}^{r,r}= p^2 + \frac{w^2}{4u} [k^2+l^2 -2 r^2] \ .
\end{equation}
Finally, by simple inversion, we get the replicon propagator 
of Eq.~(\ref{eq:RFT-repli}).

The computation of the longitudinal Fourier component of the
propagator is less straightforward. In the longitudinal-anomalous
sector the fluctuation matrix $M^{r,s}_t$ carries only a lower index
(cross-overlap), and thus only one Replica Fourier Transform is
needed. However, the lower index runs over a hierarchical tree and
when $t$ crosses the upper `passive' indices, the multiplicity may
change (sums over replica co-distances such as $r,s,t$ are
equivalent to the original sums over replica indices only if the
correct multiplicity is taken into account).  To deal with this, one
has to generalize the definition of Replica Fourier transform in
the presence of passive indices:
\begin{equation}
M^{r,s}_{\hat k}=\sum_{t=k}^{R+1}p_t^{(r,s)}
[M^{r,s}_t-M^{r,s}_{t-1}]\ ,
\end{equation} 
with, for  $r<s$, 
\begin{eqnarray}
p_t^{(r,s)}&=& p_t \quad \ \ t\le r \ ,\nonumber \\
&=&2 p_t \quad r< t\le s \ ,\nonumber \\
&=&4 p_t \quad   r<s<t\ .
\end{eqnarray}
With such a definition, the unitarity equation in the
longitudinal-anomalous sector becomes:
\begin{equation}
G^{r,s}_{\hat k}= - g_{\hat k}^r [ M^{r,s}_{\hat k}  g_{\hat k}^s + 
\sum_{t=0}^R M^{r,t}_{\hat k} \frac{\delta^{(k-1)}}{4} G^{t,s}_{\hat k}] \ ,
\end{equation}
where
\begin{eqnarray}
g_{\hat k}^r &=& G^{r,r}_{\widehat{r+1},
\widehat{r+1}} \quad k\le r+1 \ ,\nonumber \\
&=& G^{r,r}_{\widehat{r+1},\hat k} \ \ \ \quad k >  r+1\ ,
\end{eqnarray}
and
\begin{equation}
  \delta_t^k = p_t^k - p_{t+1}^k\ .
\end{equation}
This equation can be solved with the use of Gegenbauer functions
whenever $M^{r,s}_t$ depends only on $\min({r,s})$ or on $\max({r,s})$
(see \cite{DominicisKondor97} for details). The result for
$G^{r,s}_{\hat 0}$ is the one reported in Eq.~(\ref{eq:propa-la}).

Finally we note that, in the replicon geometry ($r \equiv s$),
there is a Longitudinal-Anomalous (LA) contribution
to the replicon fluctuation matrix and to the propagator. If we write 
\begin{equation} 
M^{r,r}_{u,v}= _R
M^{r,r}_{u,v}+_A M^{r,r}_{u,v}  \ ,
\end{equation} 
the LA part can be shown to be~\cite{DominicisCarlucci97}
\begin{equation} 
_A M^{r,r}_{u,v}=M^{r,r}_u+M^{r,r}_v-M^{r,r}_r \ .  
\end{equation} 
When performing a double RFT as in Eq.~(\ref{eq:RFT3}), 
the LA contribution disappears since it bears only one lower 
index and Eq.~(\ref{eq:RFT3}) is thus a purely replicon contribution.

\section{\label{app:appendix2} The case of the Heisenberg model}

Let us consider the case of the isotropic Heisenberg model with $N$ components. 
The Hamiltonian for this system reads:
\begin{equation}
{\cal H} = \int dx \ \left \{ \frac{1}{2}\sum_{i=1}^N [(\nabla\Phi_i)^2 
+ \mu \Phi_i^2]
+ \frac{g}{4!}\left(\sum_{i=1}^N \Phi_i^2 \right)^2 \right \}
\label{eq:heisenberg}
\end{equation}
In the low temperature region, for $d>2$, this system develops a spontaneous 
magnetization ${\vec M}$. By expanding Eq.~(\ref{eq:heisenberg}) around the 
mean-field solution $M^2=6 |\mu|/g$, i.e. ${\vec \Phi} = {\vec M} +{\vec \phi}$,
we get a field theory where fluctuations along the
direction of ${\vec M}$ (i.e. longitudinal modes) are massive, 
while fluctuations
transverse to it are massless (zero or Goldstone modes). 
More precisely, 
\begin{eqnarray}
{\cal H} &=& {\cal H}_{mf} + \int  dx \left \{ \frac{1}{2}[ (\nabla \phi_L)^2 + 
(\nabla {\vec\phi}_T)^2    + \frac{g}{3} M^2 \phi_L^2] \right. \nonumber \\
&+&  \left. \frac{g}{3 !}M \phi_L (\phi_L^2 + \phi_T^2) 
+ \frac{g}{4!}  (\phi_L^2+\phi_T^2)^2 \right \}\ ,
\label{eq:heisenberg2}
\end{eqnarray}
where $\phi_L$ is the longitudinal mode, 
and ${\vec \phi}_T$ the $(N-1)$-dimensional transverse mode.
The free propagators then read:
\begin{eqnarray} 
G^0_{L}(p)&=&\frac{1}{p^2+\frac{g}{3} M^2} \ ,\nonumber \\
G^0_{T}(p)&=&\frac{1}{p^2} \ .
\end{eqnarray}

\begin{figure}
\includegraphics[width=8cm,angle=0]{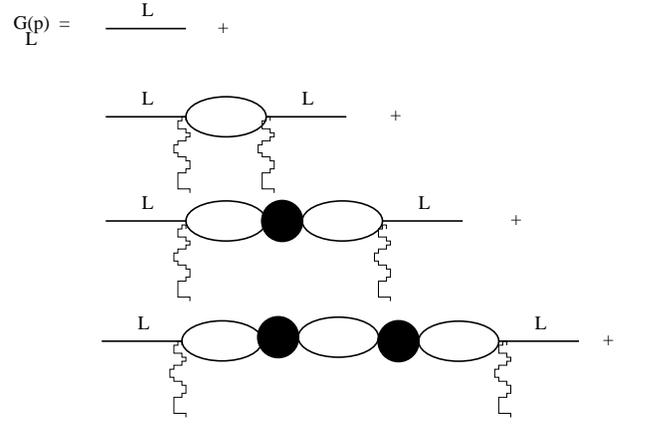}
\caption{Series expansion for the longitudinal propagator in the
Heisenberg model.}
\label{fig:prop}
\end{figure}

We now show with a simple argument that, when 
considering loop corrections, 
the longitudinal mode, which is massive at the bare level, becomes massless.
To see that, let us exhibit the series expansion of the longitudinal propagator
$G_L$ as in Fig.~\ref{fig:prop}.
Here the L-lines correspond to $G_L^0$ and the wavy lines stand for
the magnetization $M$. Since we are interested in the 
$p\to0$ limit of $G_L(p)$, all other lines 
are transverse lines, $G^0_T$. The thin vertices are couplings
$g$, the heavy vertices correspond to  the effective coupling
$\Gamma_T$ between transverse modes. We get, integrating out 
the transverse loops, 
\begin{equation}
G_L(p) \sim G_L^0(p) + M\frac{g}{3}G_L^0(p) \frac{1}{I \ p^{4-d}+\Gamma_T(p)}
M \frac{g}{3}G_L^0(p) \ ,
\label{eq:prop-resum}
\end{equation}
where $I$ is a numerical prefactor.
\begin{figure}
\includegraphics[width=8cm,angle=0]{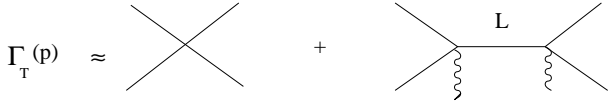}
\caption{Effective interaction between transverse modes, 
obtained by integrating out 
the longitudinal modes. }
\label{fig:int}
\end{figure}
The effective transverse interaction can be obtained by integrating out the
contribution of the longitudinal modes (see Fig.~\ref{fig:int}), yielding
\begin{equation}
\Gamma_T(p) \sim g - \frac{g^2}{3}\frac{1}{p^2+\frac{g}{3}M^2} =
\frac{g \quad p^2}{p^2+\frac{g}{3}M^2} \ .
\end{equation}
Thus, $\Gamma_T(p)$ vanishes in the infrared  limit $p\to 0$ and   
the Goldstone modes are effectively free. 
Inserting this expression in Eq.~({\ref{eq:prop-resum}), we  see that 
the massive bare behavior $1/[p^2+(g/3)M^2]$ of the 
longitudinal propagator is
changed into a massless one $1/p^{4-d}$.

This simple argument cannot be applied to the Ising spin glass.
Indeed there the Goldstone modes are the bottom of a
(transverse) band with no gap separating the massless from
the massive modes (see \cite{DominicisKondor97}). As a result,
zero modes remain coupled in the infrared, i.e., $\Gamma_T(p=0)\neq 0$,
and they develop an anomaly.


\bibliographystyle{prsty}

\bibliography{references}

\end{document}